# The Price of Removing the Scaffolding

**Maxwell wrote something about Ampère that years later would be done to him**

María Florentina Mendoza Urán

ORCID: 0009-0007-7882-6418

## Abstract

In the history of physics, the name scaffolding has been given to whatever served to raise a theory and is afterwards removed because, it is said, it is no longer needed. This article documents, from the originals, what it cost Maxwell's theory when that method was applied to it: what was removed from it, who did it, with what argument, and how over the years it had to be recovered because it was understood that it was needed after all. It also records what Maxwell himself missed in Ampère, what he calls the traces of the scaffolding: the template with which Ampère himself had given shape to his work.

## 1. Introduction

In 1873, in the third chapter of the fourth part of his Treatise, Maxwell pauses to compare two ways of setting out a discovery. The passage is in article (528), pages 162 and 163, and the running head of page 163 reads "Faraday's Scientific Method", because what he does there is to contrast two ways of working.

He speaks first of Ampère, and does so with admiration: he says that the experimental investigation by which Ampère established the laws of the mechanical action between electric currents is one of the most brilliant achievements in science, that it is perfect in form and unassailable in accuracy, and that it is summed up in a formula from which all the phenomena may be deduced, and which must always remain the cardinal formula of electro-dynamics.

And then he raises the objection:

"The method of Ampère, however, though cast into an inductive form, does not allow us to trace the formation of the ideas which guided it. We can scarcely believe that Ampère really discovered the law of action by means of the experiments which he describes. We are led to suspect, what, indeed, he tells us himself, that he discovered the law by some process which he has not shewn us, and that when he had afterwards built up a perfect demonstration he removed all traces of the scaffolding by which he had raised it."

That is where it is said what is lost, and it is not the law. The law still stands, and he has just called it unassailable. What is lost is the trace: the path by which it was reached, the formation of the ideas and what guided them. That is, the template Ampère used to give shape to his work.

Note the end of the sentence: the scaffolding by which he had raised it. By which, not beside which. A scaffolding raises nothing: it is put up to reach where one cannot reach, it is taken down, and what has been built remains the same as if it had been done another way. What Maxwell describes is something else, something that gave shape to what was raised with it. And that is why he laments its removal instead of celebrating it.

He says it, moreover, by comparison. In the following paragraph he praises Faraday for the opposite: because he shows his unsuccessful experiments as well as his successful ones, and his crude ideas as well as his developed ones, so that the reader, however inferior to him in inductive power, feels sympathy even more than admiration, and is tempted to believe that, had he had the opportunity, he too would have been a discoverer. And he closes by saying that every student should read Ampère's research as a splendid example of scientific style in the statement of a discovery, but should study Faraday for the cultivation of a scientific spirit.

He wrote that in 1873. Twelve years later they began to do the same to him.

Between 1885 and 1893, in three acts, with three dates and two signatures, a quantity was removed from his theory that he had named, defined and classified, and for which he had given a recipe for measurement. The argument was explicit and it is in writing: that it is not observed, that it complicates the matter, that it is a quantity of calculation. The textbooks were rewritten without it, and the equations taught today as Maxwell's equations are the ones that remained afterwards.

This is what was removed from him, who took it, and what went with it.

## 2. What the vector potential was in 1873

It is worth knowing what was about to be removed.

Maxwell did not invent the vector potential. The quantity came from earlier, and several had used it to calculate induced currents, as an auxiliary function within their theories. What Maxwell does with it is something else, and it is what was removed afterwards.

In the Treatise of 1873[1], the first edition and the only one Maxwell saw published, the vector potential appears with three distinct functions. All three are in the same volume, the second. The pages and articles that follow are those of that first edition, checked against the facsimile.

In article (406), page 29, he writes that it follows from the present investigation that the magnetic induction B is derived from the vector-potential A by the application of the same operator, and that the result is true within the magnet as well as outside it.

B comes from A. That is the order.

In article (590), page 214, he gives it a proper name and a recipe for measurement:

“The vector A represents in direction and magnitude the time-integral of the electromotive force which a particle placed at the point (x, y, z) would experience if the primary current were suddenly stopped.”

And he adds what he is going to call it: the Electrokinetic Momentum at that point. And he specifies that it is identical with the quantity he had investigated in article (405) under the name of the vector-potential of magnetic induction.

A proper name, and an instruction for measurement: the current is stopped suddenly, and it is measured.

And in article (592), page 216, he classifies the two of them. He says that the vector B, since it appears in a surface-integral, belongs to the category of fluxes. And that the vector A, on the other hand, belongs to the category of forces, since it appears in a line-integral.

In that same paragraph there is a sentence worth reading slowly, because it is the one usually attributed to A, and it is said of B: that in identifying that vector, which appeared as the result of a mathematical investigation, with the magnetic induction, whose properties were learned from experiments with magnets, no new fact is introduced into the theory; only a name is given to a mathematical quantity.

He says it of B. Of A he does not say it.

With those three pieces Maxwell writes, in article (598), page 221, what he labels as the equations of electromotive force (B). Three terms: the motion through the field, the time derivative of the vector potential, and the gradient of the scalar potential. And in article (599), page 222, he says in his own words that this electromotive force depends on three circumstances, and he sets them out one by one.

Note in passing which of the two quantities he declares indeterminate. In article (598) he writes that the terms involving the new quantity Ψ are introduced for the sake of giving generality to the expressions, that they disappear from the integral when it is extended round the closed circuit, and that the quantity Ψ is therefore indeterminate as far as regards the problem before him. What is indeterminate is Ψ. Of A he does not say that.

And he confirms it two articles further on. In (601), page 224, when calculating the total electromotive force round a closed circuit, he writes that the value of Ψ disappears from that integral. It is Ψ that goes; A stays.

That is what there was in 1873: a quantity with a name, with its own place in the classification, with instructions for measuring it, and with the most general law written from it.

## 3. What Heaviside had before removing it

It is worth looking at what Heaviside had in his hands in the months before January 1885, because it was not what is usually assumed.

On 25 November 1882, in The Electrician, he publishes the second section of his article 24. It is entitled “The Potentials of Scalars and Vectors”, and it begins on page 201 of the first volume of his Electrical Papers[2]. There he introduces, in his own words, the concept that the potential of a vector is a vector, and warns that some explanation is needed to make it intelligible. And he states that the magnetic force is the vector-potential of the curl of the current.

That same article has six sections, and three are about potentials: “The Potentials of Scalars and Vectors”; “The characteristic equation of a potential, and its solution”; and “Relations of Curl and Potential, direct and inverse. Scalar Potential of a Vector”.

And in article 25, “The Energy of the Electric Current”, a whole section bears the title “Probable Localization of the Energy. Division of any Vector into a Circuital and a Divergent Vector”.

And in article 26, “Some Electrostatic and Magnetic Relations”, published in The Electrician between April and June 1883, the seventh section is entitled, literally, “Complete Scheme of Potentials”. It came out on 28 April 1883 and begins on page 262 of the volume.

What is there is not a passing mention. Heaviside builds a ladder. He sets up a series of quantities in which each vector is the vector-potential of the next and each scalar is the divergence of the preceding vector, alternating scalars and vectors. And he applies it to the magnetic field: he separates the magnetic force B, in his notation, into two parts, $B_1$ without convergence and $B_2$ without curl; he calls $C_1$ the curl of $B_1$, which is the current-density, and $C_2$ the divergence of $B_2$, which is the density of magnetism. And then he writes, on page 263, that $A_1$ is the vector-potential of $C_1$ and $A_2$ the scalar potential of $C_2$. He adds Z above and D below, and sets out a table of five columns, Z, A, B, C, D, marking under each whether it is a scalar or a vector.

And he gives the relations in both directions. Downwards: A is the curl of Z, and B is the curl of A. Upwards: Z is the potential of B, and A is the potential of C. And he closes by saying that Z is clearly the vector-potential of B, and B the vector-potential of D.

That is: in the spring of 1883, Heaviside did not merely handle the vector potential. He had a scheme of his own, with his notation, his letter A, his table and his title.

Twenty months later the series of January 1885 begins. And in the fundamental equations he establishes in it, none of these quantities appears.

## 4. The removal: the act, the name and the judgment

Twelve years after the Treatise, the piece is removed. And the removal has three moments with three dates: the act, the name and the judgment. They are told here grouped by author.

**Heaviside**

In 1885, in the series "Electromagnetic Induction and Its Propagation"[3], which begins to appear in The Electrician on 3 January—the date is given by himself, in a footnote on page 67 of his Electromagnetic Theory—Oliver Heaviside takes the potentials out of the fundamental equations of Maxwell's theory and rewrites it in four equations with four fields: the ones taught today as Maxwell's equations.

It is worth seeing how he does it, because he does not argue it: he makes it disappear by construction.

The first section of the series, that of 3 January 1885, is entitled "Rough Sketch of Maxwell's Theory", and begins on page 429 of the first volume of his Electrical Papers. There he sets up the scheme from scratch, with the conductivity, the capacity and the permeability, and in all of it the vector potential does not appear.

The elimination comes in the fourth section, on page 448. When establishing the second relation between the electric force and the magnetic force, Heaviside warns that in his equation E is the electric force of induction only, not the actual electric force, and that there may in addition be electrostatic force. And then he writes that the electrostatic force is polar, that it is derived from a scalar potential, that if this be P the force is minus the gradient of P, and that, since the curl of a gradient is zero, that polar force may simply be included in E. And he adds that the same holds for H in the previous equation.

That is: the potential is not removed with an argument. It is absorbed into the field, and disappears on taking the curl.

From there come his two equations, the two circuital laws, written with the electric force, the magnetic force and the impressed forces. And he closes them with a sentence: that we now have a dynamically complete system.

It is worth specifying what goes out and what stays, because they are not the same thing.

What goes out of the fundamental equations is the potential. But further on, in that same series, A reappears: on page 467 Heaviside defines Z as the vector-potential of the magnetic current and writes, in his equation (67a), "If A be Maxwell's vector-potential of the electric current". And on page 468 he writes equation (73a), E equal to minus A-dot minus the gradient of P, and notes beside it: "(73a) is Maxwell's equation".

That is: A does not disappear from his work. What it loses is its rank. It ceases to be that from which one starts, and becomes something to be resorted to when convenient.

With the scalar potential he does something else, and it is more forceful. In April 1889, in the Philosophical Magazine[4], he writes that the divergence of A is of no importance and that introducing Ψ is only an annoying complication. And on the following page he entitles a heading "Complete Solution in the Case of Steady Rectilinear Motion. Physical Inanity of Ψ".

A he demotes. Ψ he expels.

The judgment comes in 1893. In "Electromagnetic Theory"[5], volume one, published in London that year, he writes on page 46 that the method by which Maxwell deduced the electric force of motion is substantially the same in principle; that he, however, makes use of an auxiliary function, the vector-potential of the electric current, and that this rather complicates the matter, especially as regards the physical meaning of the process; and that it is always desirable when possible to keep as near as one can to first principles.

In the preface of that same volume, dated 16 December 1893, he says it in another way: that the forces and fluxes are, in his exposition, the objects of immediate attention, instead of the potential functions, which are such powerful aids to obscuring and complicating the subject and to hiding from view useful and sometimes important relations. And a few pages later he declares the substitution in so many words: that among the differences between his exposition and Maxwell's, and among those he calls changes of form only, is the substitution of the second circuital law for Maxwell's equation of electromotive force involving the potentials.

That equation of electromotive force involving the potentials is the one Maxwell labels (B) in article (598). The three terms.

But where he says it in full is in the second chapter, paragraph 65, page 69. There he tells what he found and what he did with it. He writes that the mathematicians, instead of putting into symbols what Faraday had recognized, worked in a more indirect way, and expressed it by means of an equation of electromotive force containing a function called the vector potential of the current, and another potential, the electrostatic, working together not altogether in the most harmoniously intelligible manner; and then he opens a dash and translates: "in plain English, muddling one another".

And next he declares the act:

"Finding these equations of propagation containing the two potentials unmanageable, and also not sufficiently comprehensive, I was obliged to dispense with them; and, going back to first principles, introduced what I term the second circuital law as a fundamental equation, the natural companion to the first."

And in the next sentence, on that same page, he says what is gained by the change: that it makes it possible to simplify and clarify considerably the treatment of general questions, whilst bringing to light interesting relations which were formerly hidden from view "by the intervention of the vector potential A, and its parasites J and Ψ".

Parasites. He does not say that they are superfluous; he says that they live at the expense of something.

There is also a technical reason, and he gives it a few pages earlier, in paragraph 63, page 66: that this method, or the equivalent methods employing potentials, is quite inadequate for the treatment of electromagnetic waves, and is then usually of a quite unpractical nature.

And it is worth recording in what spirit he does all this, because he writes it himself on page 68. He says that there are spots in the sun, and that he sees no good reason why the many defects of Maxwell's treatise should be ignored; that it is most objectionable to stereotype the work of a great man, apparently merely because it was such a great advance and because of the great respect it induces. On the same page he acknowledges that Maxwell, although he did not himself explicitly represent the convection current, which he calls a notable oversight, insisted strongly on the circuital nature of the electric current, and that he would have seen the oversight as soon as it was suggested to him.

**Hertz**

In 1890, dated 19 March, Heinrich Hertz published in the Göttinger Nachrichten a paper entitled "On the fundamental equations of electrodynamics for bodies at rest"[*6], which appeared that same year in Wiedemann's Annalen, volume 40, pages 577 to 624.

In it he confirms what Heaviside had done: he writes that Mr. Oliver Heaviside has been working in the same direction since 1885, that the concepts which Heaviside removes from Maxwell's equations are the same ones that he removes, that the simplest form those equations take is, apart from secondary matters, the same one at which he arrives, and that in this respect the priority belongs to Heaviside. And he adds in a footnote that those equations are to be found in the Philosophical Magazine of February 1888, that reference is made there to earlier works in The Electrician of 1885, and that this source has been inaccessible to him.

Hertz opens by acknowledging the value of what he is going to touch: he says that the system of concepts and formulae with which Maxwell represented electromagnetic phenomena is, in its possible development, richer and more comprehensive than any other devised for the same purpose, and that precisely for that reason it would be desirable to perfect it in its form as well, so that the construction of the system lets its logical foundations be recognized transparently, every non-essential concept is removed from it, and the relations between the essential concepts are reduced to their simplest form.

And next he says from where he reads Maxwell. He writes that Maxwell's own presentation does not mark, in that respect, the attainable goal, and that it frequently wavers between the conceptions Maxwell found and those at which he arrived. And he explains it: that Maxwell starts from the assumption of unmediated forces at a distance, investigates the laws according to which, under the influence of such forces at a distance, the hypothetical polarizations of the dielectric ether change, and ends with the assertion that these polarizations really do change in that way, without forces at a distance being in truth their cause.

And he says what is superfluous. He writes, on page 209, that this path leaves in the formulae a number of superfluous concepts, rudimentary so to speak, which had their proper meaning only in the old theory of unmediated action at a distance. As rudimentary concepts of a physical nature he names the dielectric displacement in the free ether, distinct from the electric force that produces it, and the ratio between the two, the dielectric constant of the ether.

And there is the name. After mentioning, as a rudimentary phenomenon of a mathematical nature, the predominance of the vector-potential in the fundamental equations, he writes:

"In the construction of the new theories the potentials served as scaffolding, in that through their introduction the forces at a distance, appearing discontinuously at individual points, were replaced by quantities which at every point of space are conditioned only by the states of the neighbouring points. But after we have learned to regard the forces themselves as quantities of the latter kind, their replacement by potentials has a purpose only if a mathematical advantage is thereby attained."*

And he goes on: that such an advantage does not seem to him to be associated with the introduction of the vector-potential into the fundamental equations, in which in any case one may expect to find relations between quantities of physical observation, not between quantities of calculation.

It is worth seeing how he does it, because he does not stop at the argument: he carries it into the equations by construction, and he announces it. On page 210 he divides the work into two parts. In the first he gives the fundamental concepts and the formulae that connect them; in the second he derives the phenomena from those formulae.

In the first part, on page 214, he writes the equations of the ether with the electric and the magnetic force only: two of curl and two of divergence. He does not derive them. He says that the forces in the ether are connected according to those equations, and adds that, once they have been found, it no longer seems appropriate to derive them from conjectures about the electric and magnetic constitution of the ether, as would correspond to the historical course, but that it is more appropriate to attach to these equations the conjectures about that constitution. On page 224 he establishes that in the ether polarization and force coincide, and with that the two physical concepts he had declared rudimentary disappear. In the whole of this first part, the potential does not appear.

In the second part the potentials return, one by one, each time a particular case allows it. In electrostatics, on page 237, since the force has no curl, he writes that the forces therefore possess a potential. In stationary currents, on page 241, that the forces in the interior of the conductors can be represented as derivatives of a function, the potential. For the magnetic forces of those currents, on pages 242 and 243, he introduces, in his own words, as auxiliary quantities the so-called components of the vector potential, three integrals of the current. And for the work between closed currents and for induction, on pages 246 and 249, Neumann's potential of one circuit on another.

It is the same thing Heaviside did. The potential does not disappear from Hertz's work: it ceases to be in the fundamental equations and becomes a tool to be resorted to in particular cases. And in the case of the scalar the direction changes as well: it is no longer that from which the forces are obtained, but something the forces possess when they have no curl.

There is, however, a difference between the two, and it lies in how they name it when it returns. Both name Maxwell when they explain what they remove. But Heaviside, when A returns, attributes it: he writes "If A be Maxwell's vector-potential of the electric current", and notes beside his equation that it is Maxwell's equation. Hertz, on page 243, introduces it as the so-called components of the vector potential.

## 5. The scaffolding dragged a law along with it

What was removed was not a notation. It was a law, and it is worth placing the texts side by side to see how large it was.

Heaviside deduces one term: the electric force of motion. Maxwell deduces three, and says where each one comes from. And Maxwell's equations of propagation, which contained the two potentials, Heaviside calls unmanageable and not sufficiently comprehensive: a method that delivers three terms, against his own, which delivers one.

And note what he calls the piece: an auxiliary function. And what goes with it, J and Ψ, its parasites. In the Treatise that piece has a proper name, Electrokinetic Momentum, article (590); it has a recipe for measurement on that same page; and it has a place in the classification, article (592), in the category of forces.

There is no dispute about the calculation. There is a demotion in category.

Hertz performs the same operation with a different argument. He does not say it is wrong: he says that in the fundamental equations one may expect to find relations between quantities of physical observation, not between quantities of calculation. The criterion is that of observability.

So what was lost was not a result. It was the form in which that result was written: a general law with three terms, deduced from a quantity with a name and a recipe for measurement, and which said where each part of the electromotive force came from.

And this is where the word oblivion describes badly what happened. Oblivion is what happens to something nobody touched. Here there are three acts with a date and a signature: 1885, when Heaviside takes them out of the fundamental equations; 1890, when Hertz gives the reason; 1893, when Heaviside publishes the judgment. Oblivion comes afterwards, and is a consequence.

And that it was not an oversight is proved by what happens in 1888. In a postscript dated 18 October, entitled “On the Metaphysical Nature of the Propagation of the Potentials”, Heaviside writes that, according to the way of regarding electromagnetic quantities that he has consistently carried out since January 1885, the question of the propagation of the electric potential Ψ and of the vector potential A does not present itself as a question for discussion, and that, when it is raised, it turns out to be of a metaphysical nature.

In a letter from Sir William Thomson, which Heaviside publishes with his permission, the latter contradicts him. Thomson maintains that the speed of propagation of the electric potential is not a merely metaphysical question, proposes an experiment to measure it, and adds that neither is that of the vector potential, which is nothing but the speed of propagation of the electromagnetic force.

Heaviside publishes it at the beginning of article 47, and stands his ground: he writes that he is accustomed to picturing the electric and magnetic forces because they give the most direct representation of the state of the medium, which is, he says, the true physical subject of propagation, and that reaching that state through the vector potential requires complex operations. With Kelvin facing him, he does not move.

And it is worth making clear, to finish, that this loss produced no error. What remained standing after 1885 works: the four equations written with the fields alone are complete for classical electrodynamics, and everything that came afterwards was built with them.

But that is not an argument against; it is exactly what was to be expected. A template is thrown away and the piece still stands, because the form stayed inside. What is lost in throwing it away is not the result: it is the trace that explained why it came out that way.

## 6. The criterion

A criterion is therefore needed, and it comes from Maxwell’s own passage. Not to ask whether the work stands without what was removed, because a well-made work always stands, and he himself said that Ampère’s law was unassailable. To ask whether what was removed is still explained after removing it. If it is, it was scaffolding. If not, it gave shape to what was raised with it, and its removal takes the trace with it.

Removing a scaffolding and saying that it has been removed are not the same operation. The second leaves a trace, and the trace is dated.

One observation remains about the figure itself. The reason that was given was of a particular kind: that it is not observed, that it is not the true mechanism, that it is a quantity of calculation. It is a reasonable criterion, and in its time it was fruitful. But it measures something different from what the metaphor suggests. That something is not observed with the instruments of a given year says nothing about whether it left its form inside the work. With the potential, it took seventy-four years to find the way to observe it.

## 7. What was removed as useless was in the end used as a template

It stayed that way for decades. The textbooks were rewritten without the potential, and the piece remained inside the equations without anyone claiming its name.

Until it was needed.

In 1959, in Physical Review, Yakir Aharonov and David Bohm published “Significance of Electromagnetic Potentials in the Quantum Theory”[7]. They maintain there that, contrary to the conclusions of classical mechanics, there exist effects of potentials on charged particles even in the region where all the fields, and therefore all the forces on the particles, vanish. And they propose experiments to test it.

The setup is the one Feynman describes in his lecture of 1962, section 15-5[8]: a beam of electrons that is split and passes on both sides of a solenoid, through a region where the magnetic field is zero. The interference pattern shifts. If the fields were enough, nothing could happen there. It happens.

Robert G. Chambers measured it the following year, in 1960, in Physical Review Letters[9]. Akira Tonomura and his team confirmed it in 1982 by electron holography[10] and again in 1986 with the magnetic field completely shielded by a superconductor[11]. There is also an earlier theoretical antecedent, by Werner Ehrenberg and Raymond Siday[12], published in 1949, which went unnoticed for ten years.

Seventy-four years, from 1885 to 1959, to find the way to observe what had been removed as unobservable.

And there is one last thing in that section of Feynman’s that is worth recording, because it tells the same story this article documents. He writes that the theory was known from the beginning of quantum mechanics, in 1926—that is, from the Schrödinger equation, which is the date he takes as the beginning—; that the fact that the vector potential appears in that equation was obvious from the day it was written; and that, one after another, those who tried to replace it by the magnetic field observed that it could not be done in any easy way. And he goes on: that, in spite of this, people repeatedly said that the vector potential had no direct physical significance, that only the magnetic and electric fields are the right ones, even in quantum mechanics.

And he closes: that it seems strange in retrospect that no one thought of discussing this experiment until 1956, when Bohm and Aharonov first suggested it—the date is Feynman’s; the published paper is from 1959—; that the implication was there all the time, but no one paid attention to it; and that it is interesting that something like this can be around for thirty years but, because of certain prejudices about what is and what is not significant, continues to be ignored.

Oblivion, told by someone who saw it from inside, and in class.

And on the status of the piece, in section 15-4 he puts it in writing:

"It turns out, however, that there are phenomena involving quantum mechanics which show that the field A is in fact a 'real' field in the sense we have defined it."

And in the following section, once he has described the experiment, he says it plainly:

"In our sense then, the A-field is 'real.'"

And in the same chapter he puts in writing why it had to be recovered. He says that it is precisely because momentum and energy play a central role in quantum mechanics that the potentials provide the most direct way of introducing electromagnetic effects into quantum descriptions; and that those who tried, one after another, to replace the vector potential by the magnetic field observed that it could not be done in any easy way.

Hertz had removed it from the fundamental equations in 1890 because, he wrote, in them one may expect to find relations between quantities of physical observation. The quantity was observed. And by then it had already returned to the equations, because it was needed.

And with the last sentence of that same section, Feynman states where all this was heading:

"E and B are slowly disappearing from the modern expression of physical laws; they are being replaced by A and ϕ."

It is an observation of his on how the equations are written today in quantum field theory, and it is recorded here for its symmetry with Hertz's. What this work documents is something else: what was removed in 1885, with what argument, and what went with it.

In 1900, in the Archives Néerlandaises, Emil Wiechert writes the equations with the vector potential, which he designates by Γ, and explains why he chooses it: because that system, he says, is in many cases more advantageous and adheres more closely to Maxwell[13]. He names him. And on page 561 he writes the potentials evaluated at the retarded time, the ones physics ended up calling the Liénard-Wiechert potentials[14].

No one gave the piece back to Maxwell, because no one had noticed that it had been taken from him. It came back because it was needed. The potential, through Wiechert, who drew its retardation from it while naming Maxwell. And the law, through Lorentz[15], who worked on Maxwell's to draw out his own.

Both worked on what had been removed from Maxwell, and both were able to give a name to what came out of it: the Liénard-Wiechert retarded potentials, the Lorentz force.

And yet, to this day, in spite of having been used as a template for the work of others, neither the vector potential nor its law has recovered Maxwell's name.

* Note on the translations. All quotations in this article come from the originals, checked against facsimile. Those from Maxwell, Heaviside, Feynman, and Aharonov and Bohm are given in their original English; in those from Maxwell's Treatise, the vectors, printed there in German capitals, are rendered with the Roman letters used throughout this article. Those from Hertz and Wiechert have been translated literally by the author directly from the German original, respecting the usage of their place and period, without adapting or softening any expression. Those from Hertz have not been taken from the authorized English translation by Daniel Evan Jones, "Electric Waves", Macmillan and Co., London, 1893, because it was found to differ from the original, beginning with the title of the paper itself. For the same fidelity, constructions that sound unnatural in English have been kept, such as "in that through their introduction" in the passage on the scaffolding, which renders the "indem durch ihre Einführung" of the original.